\documentclass[twocolumn,showpacs,preprintnumbers,amsmath,amssymb,prl,superscriptaddress]{revtex4}

\usepackage{amsfonts}
\usepackage{graphicx}
\usepackage{dcolumn}
\usepackage{bm}
\usepackage{color}

\begin{document}

\preprint{APS/123-QED}

\title{Thermoelectric response near a quantum critical point of $\beta$-YbAlB$_4$ and YbRh$_2$Si$_2$\\: A comparative study}

\author{Y. Machida}
\author{K. Tomokuni}
\author{C. Ogura}
\author{K. Izawa}
\affiliation{Department of Physics, Tokyo Institute of Technology, Meguro 152-8551, Japan}

\author{K. Kuga}
\author{S. Nakatsuji}
\affiliation{Institute for Solid State Physics, University of Tokyo, Kashiwa 277-8581, Japan}

\author{G. Lapertot}
\author{G. Knebel}
\author{J.-P. Brison}
\author{J. Flouquet}
\affiliation{INAC, SPSMS, CEA Grenoble, 38054 Grenoble, France}

\date{\today}

\begin{abstract}
The thermoelectric coefficients have been measured on the Yb-based heavy fermion compounds
$\beta$-YbAlB$_4$ and YbRh$_2$Si$_2$ down to a very low temperature.
We observe a striking difference in the behavior of the Seebeck coefficient, 
$S$ in the vicinity of the Quantum Critical Point (QCP) in the two systems.  
As the critical field is approached, $S/T$ enhances in $\beta$-YbAlB$_4$ but is drastically reduced in YbRh$_2$Si$_2$. 
While in the former system, the ratio of thermopower-to-specific heat remains constant, it drastically drops near the QCP in YbRh$_2$Si$_2$. 
In both systems, on the other hand, the Nernst coefficient shows a diverging behavior near the QCP. 
The results provide a new window to the way various energy scales of the system behave and eventually vanish near a QCP.
% On approaching their quantum critical points (QCPs), we find a highly contrasting behavior of the Seebeck coefficient $S$;
% $-S/T$ for $\beta$-YbAlB$_4$ exhibits a diverging enhancement in contrast with a considerable decrease for YbRh$_2$Si$_2$.
% Consequently, a semi-universal relation between the Seebeck coefficient and the electronic specific heat 
% is hold in $\beta$-YbAlB$_4$, while a significant deviation is found in YbRh$_2$Si$_2$,
% suggesting the distinct nature of the quantum criticality
% between the two systems.
% On the other hand, a remarkable increase of the Nernst coefficient is similarly observed in both systems as an indication of vanishing of energy scale of the Fermi liquid
% towards the QCPs.
% A comparative study via a dimensionless ratio of the Seebeck coefficient and the electronic specific heat 
% clearly reveals a significant deviation from the semi-universal value in YbRh$_2$Si$_2$, 
\end{abstract}

\pacs{71.27.+a, 72.15.Jf, 74.40.Kb}

\maketitle
Heavy fermion compounds have been intensively studied for their ability to use pressure or magnetic field 
to access a quantum critical point (QCP) from a magnetic phase or a Fermi liquid (FL) state~\cite{gegenwart_review}.
One consequence of the QCP is an emergence of non-Fermi liquid (NFL) behavior revealing strong deviations from the FL. 
% One consequence of the QCP is the vanishing characteristic energy scale of the Fermi liquid. 
The behavior has been characterized by a set of exponents of the temperature dependence of the physical properties, such as the specific heat, the resistivity, and the susceptibility
in the framework of the spin fluctuation theories~\cite{moriya,hertz,millis}. 
So far, these theories have been succeeded to describe the NFL properties of a number of materials~\cite{knafo}.
% So far, this behavior has been indicated mainly by the divergence of the electronic specific heat [1]. 
Recently, however, there has been found a few heavy fermion materials, 
e.g., CeCu$_{1-x}$Au$_x$~\cite{lohneysen}, YbRh$_2$Si$_2$~\cite{gegenwart}, and $\beta$-YbAlB$_4$~\cite{nakatsuji}, 
exhibiting the NFL behavior in the neighborhood of the QCP 
for which the standard theory is insufficient.
% There exists, however, a few examples of heavy fermion systems which do not obey the conventional framework.
In spite of  the considerable efforts, both experimental and theoretical, 
up to date nature of their unconventional criticality has still been debated.
% In some cases, however, the dominant nuclear contributions at low temperatures prevent the detailed studies on the QCP by means of the specific heat. 
One ingredient preventing a clarification of the QCPs is due to a difficulty of precise determination of exponents for the NFL behaviors.
A nuclear contribution to the specific heat, for example, tends to mask an intrinsic behavior at low temperatures,
and necessarily leads to ambiguity.
% because of their high sensitivity to impurities and/or necessity of experiments under extreme conditions, e.g. low temperature, high magnetic field and high pressure.
% One major factor preventing identification of the unconventional QCP comes from lack of systematic study

Recently, new insight on the QCP via the thermoelectric coefficients, e.g., the Seebeck and Nernst coefficients, has been examined~\cite{izawa,hartmann}.
An advantage of these quantities is that only itinerant quasiparticle (QP) excitations are measured,
%  and high sensitivity to the low-energy excitations at low temperatures,
so thus they would be complementary to the specific heat: a measure of both itinerant and localized excitations.
Moreover, in contrast to the specific heat, the thermoelectric response is not determined 
by the number of itinerant electrons present in the system and its magnitude is directly set by their relevant energy scale.
In this paper, we report on a comparative study of the thermoelectric response near the QCP of the two representative 
Yb-based unconventional quantum critical materials,
$\beta$-YbAlB$_4$ and YbRh$_2$Si$_2$.
% which allows us to discriminate the nature of their QCPs.
We find that while the Nernst response behaves in a similar manner in the two systems, the Seebeck coefficient 
and its relation to the specific heat are quite distinct in the two cases, allowing discrimination of the nature of their quantum criticality.

$\beta$-YbAlB$_4$ is the first Yb-based heavy fermion
superconductor ($T_c$ = 80 mK)~\cite{nakatsuji}.
The striking feature of this system is an emergence of the NFL behaviors
under zero field at ambient pressure.
This implies that the system is located near the QCP without the external tuning parameter~\cite{matsumoto}.
Coexistence of valence fluctuation and the NFL behavior is another feature of this material~\cite{okawa}.
% Recently, the field quantum criticality has been confirmed by a scaling of Free energy.
YbRh$_2$Si$_2$ is a heavy fermion
compound showing the NFL properties above the antiferromagnetic (AF)
transition at $T_{\rm N}=$ 70 mK~\cite{gegenwart,custers}.
Significantly, $T_{\rm N}$ is found to be
continuously suppressed with a small magnetic field ($B_{\rm c}$ = 0.66 T $\parallel c$-axis) to zero
accessing the field-induced QCP.
An existence of additional energy scale $T^*$ discussed as the crossover line of the
 Kondo destruction is a puzzling feature of YbRh$_2$Si$_2$~\cite{paschen}.
A key characteristics of both systems, which goes beyond the standard theory, is a diverging enhancement of the 
effective mass and the uniform susceptibility towards the QCP~\cite{gegenwart,custers,nakatsuji}.
Stimulated by these striking similarity, the theoretical efforts have been devoted to 
understand the unconventional quantum criticality of the two systems in a unified way~\cite{custers2,imada,watanabe},
but no consensus has been yet reached.
% The implications of these observations to the unconventional quantum criticality has been intensively studied,
% but yet to be clarified~\cite{imada,watanabe}.

Single crystals of $\beta$-YbAlB$_{4}$ and YbRh$_2$Si$_2$
were prepared using the aluminum and indium flux, respectively~\cite{robin,knebel}.
We used two crystals with different qualities for $\beta$-YbAlB$_{4}$ 
(the residual resistivity ratio ($RRR$) is 130 and 270 for sample $\sharp$ 1 and $\sharp$ 2, respectively),
and one crystal ($RRR\sim$ 90) for YbRh$_2$Si$_2$.
The thermoelectric coefficients were measured
by employing a steady-state method in a dilution refrigerator. 
The heat current $q$ was injected parallel to
the $ab$-plane, and the magnetic field
was applied parallel to the $c$-axis on the sample with a size of
typically $0.5 \times 0.1 \times 0.005$ mm$^3$ for $\beta$-YbAlB$_{4}$ and $1.6 \times 2.2 \times 0.05$ mm$^3$ for YbRh$_2$Si$_2$, respectively.
The thermal contacts with resistance of $\leq$ 10 m$\Omega$ at room temperature
were made by using a spot welding technique.
The same contacts were used to measure the resistivity by a standard four-contact method.

\begin{figure}[t]
\begin{center}
\includegraphics[scale =0.45]{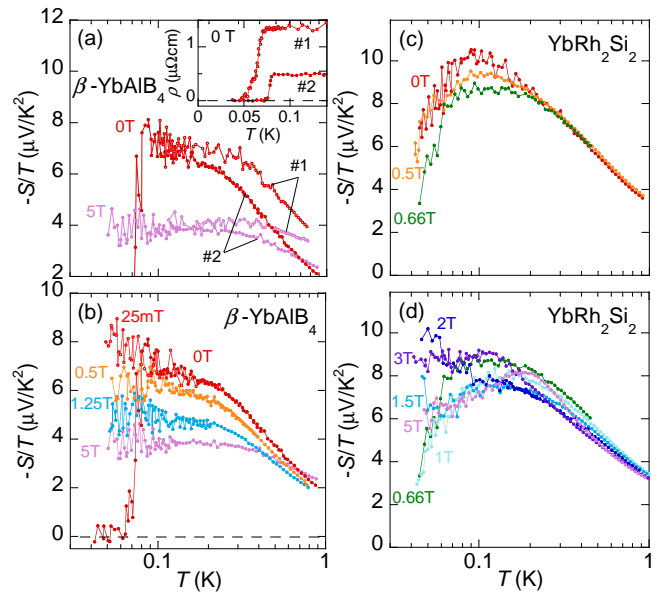}
\end{center}
\vspace{-0.5cm}
\caption{\label{fig.1} (color online). (a) Temperature dependence of the Seebeck coefficient $-S(T)/T$ 
for two crystals of $\beta$-YbAlB$_{4}$ (sample $\sharp$1 and $\sharp$2) under zero field and 5 T. 
Inset: zero-field resistivity vs $T$ plot for $\sharp1$ and $\sharp2$.
(b) $-S(T)/T$ curves under several fields for $\sharp2$. 
The $-S(T)/T$ curves for YbRh$_2$Si$_2$ (c) below and (d) above the critical field of 0.66 T, respectively.}
\end{figure}

% Next, let us discuss about the sample dependence of $S/T$.
% First, we demonstrate that the Seebeck coefficient is a sensitive probe of the low-energy excitations
% by presenting the sample-independent behavior at low temperatures. 
Figure~\ref{fig.1}(a) shows the temperature dependence of the Seebeck coefficient divided by temperature $-S(T)/T$ under zero field and 5 T
for the two different samples of $\beta$-YbAlB$_{4}$ (sample $\sharp$1 and $\sharp$2).
What is remarkable is that the $-S(T)/T$ curves for sample $\sharp$1 and $\sharp$2 converge at $T<$ 0.15 K under each field,
even though the residual resistivity is different by a factor of 2.5 (the inset of Fig.~\ref{fig.1}(a)).
This indicates that the Seebeck coefficient takes an identical value 
at sufficiently low temperature, irrespective of the sample quality.
Using the Boltzmann equation, the Seebeck coefficient is expressed as
$S/T=-(\pi^2/3)(k_{\rm B}^2/e)(\partial \ln\sigma(\epsilon)/\partial \epsilon$)$_{\epsilon=\epsilon_{\rm F}}$ ($\sigma$: the electrical conductivity).
In the simple case of a free electron gas, the term $(\partial \ln\sigma(\epsilon)/\partial \epsilon$)$_{\epsilon=\epsilon_{\rm F}}$ is approximated by 
the energy derivative of the scattering time,
$(\partial \ln\tau(\epsilon)/\partial \epsilon$)$_{\epsilon=\epsilon_{\rm F}}$.
Moreover, in the zero-temperature limit, it can be further simplified as~\cite{behnia_seebeck} 
\begin{equation}
\frac{S}{T}\simeq-\frac{\pi^2}{3}\frac{k_{\rm B}^2}{e}\frac{N(\epsilon_{\rm F})}{n}\biggl{(}1+\frac{2\zeta}{3}\biggr{)},
\end{equation}
by expressing $\tau(\epsilon)=\tau_0\epsilon^{\zeta}$, 
where $N$, $\epsilon_{\rm F}$, and $n$ are the density of states, the Fermi energy, and the carrier concentration, respectively.
Therefore, the Seebeck coefficient is proportional to the density of states per electron as $T\rightarrow$ 0.
% examining the sample dependence of the Seebeck coefficient in the low-temperature limit.
% Note that the sign of $S$ is positive for Ce-based compounds and negative for Yb-.
% The sign of $S$ is controlled by the carrier type: positive for holes, negative for electrons.
% From Eq. (3), one expects that th Seebeck coefficent depends on the delocalized quasiparticle density of states $N(\epsilon_{\rm F})$ and 
% the exponent of enegy dependence of the relaxation time $\zeta$, and does not  depend on the $\tau$.
% \textcolor{red}{This indicates that the Seebeck coefficient is a good measure of the low-energy excitations 
% at sufficiently low temperature irrespective of the sample quality, as
% expected from Eq. (1).}
% In the FL state, $N(\epsilon_{\rm F})$ is temperature independent, and thus $S/T$ becomes constant.
% reliability of this estimation, valid as $T\rightarrow$ 0.
% indicating that the thermopower is sample-independent in the low-temperature limit.
This fact gives rise to an intriguing aspect of $S/T$ that it is interrelated with the electronic specific heat coefficient $C/T=\gamma_0$
via the dimensionless ratio $q=(S/T\gamma_0)N_{\rm AV}e\approx \pm1$ ($N_{\rm AV}$: the Avogadro number),
since both quantities are proportional to the density of states.
It has been shown that this semi-universality is held within logarithmic accuracy in a wide variety of metals including the heavy fermions~\cite{behnia_seebeck}.
However, an insight into this ratio in the vicinity of the QCP remains an issue of debate.
% We will address this point for $\beta$-YbAlB$_{4}$ and YbRh$_2$Si$_2$ below.
% We will get insight into the behavior of $q$ in the vicinity of the QCP for $\beta$-YbAlB$_{4}$ and YbRh$_2$Si$_2$ later.
% This issue is addressed below.
% Moreover, this fact indicates that the thermopower is a good measure of the densities of state 
% $N(\varepsilon_{\rm F})$ at $T \rightarrow 0$, 
% which allows us to examine the dimensionless ratio $q$ in the temperature range of $T <$ 0.15 K.

With the above considerations in mind, let us look at variations of $S(T)/T$ under fields.
Figure~\ref{fig.1} shows the $-S(T)/T$  vs $T$ plots under
several magnetic fields for (b) $\beta$-YbAlB$_{4}$ (sample $\sharp$2) and (c), (d) YbRh$_2$Si$_2$, respectively.
In both systems, the Seebeck coefficient takes a negative sign as observed in other Yb-based heavy fermion compounds~\cite{behnia_seebeck}.
First, we focus on the results of $\beta$-YbAlB$_{4}$.
Under zero field, $-S(T)/T$ exhibits a logarithmic increase down to 0.2 K and reaches a large value $-S/T \sim$ 6 $\mu$V/K$^2$, being
two order of magnitude larger than the values for simple metals such as Cu ($S/T\sim$ $-$30 nV/K$^2$)~\cite{behnia_seebeck}.
The large $S/T$ was also found in other heavy-fermion compounds~\cite{behnia_seebeck,izawa,hartmann} and attributed to
a large effective mass. 
% Indeed, the dimensionless ratio ($q=\frac{SN_Ae}{C}$, $N_A$ and $e$ represent the Avogadro
% number and the charge of electron, respectively) linking the thermoelectric power and the specific heat $C$ is found to be of the order of
% unity in a wide range of systems~\cite{behnia_seebeck}.
On further decreasing the temperature, $-S(T)/T$ shows a steep increase 
followed by a sudden drop at 80 mK due to the superconducting transition.
By contrast, under the field of 25 mT which is larger than the upper critical field~\cite{kuga} but very close to the QCP, 
the drop disappears and $-S(T)/T$ is found to continuously increase down to
the lowest temperature.
% Note that this behavior is incompatible with the Fermi liquid behavior ($S(T)/T \sim$ const.).
By increasing the field $B\geq0.5$ T, the low-temperature increase is suppressed and $-S(T)/T$ approaches constant,
indicating that the system becomes the FL.

Next, we turn to the results of YbRh$_2$Si$_2$.
The $-S(T)/T$ curves taken under the fields below and above the critical field ($B_{\rm c}$ = 0.66 T) are shown in Fig.~\ref{fig.1}(c) and \ref{fig.1}(d), respectively.
Under zero field, $-S(T)/T$ shows a logarithmic increase at high temperature $T>0.2$ K, the same as $\beta$-YbAlB$_{4}$.
However, upon cooling, $-S(T)/T$ peaks around 0.1 K which is slightly above $T_{\rm N}$, and then turns to decrease.
By applying the fields up to 1.0 T, the peaks broaden and the decrease of $-S(T)/T$ becomes more remarkable.
It should be emphasized that, in contrast to $\beta$-YbAlB$_{4}$, we find a considerable decrease of $-S(T)/T$ under the critical field,
while the logarithmic behavior is still observed at $T>$ 0.2 K.
% Similar logarithmic divergences of $S/T$ were also observed in MnSi and YbAgGe in the vicinity of their QCPs,
% but in the limited temperature range down to $\sim$ 5 K for the former and $\sim$ 0.35 K for the latter,
% and claimed to be a signature of the quantum criticality.
Theoretically, such logarithmic variation was predicted both in the spin-density-wave (SDW)~\cite{kotliar} and the Kondo breakdown scenarios~\cite{pepin}.
% In the framework of a two-dimensional spin-density wave scenario, calculations also predict that $S/T\propto-\ln T$.
Although the experimental results agree with these predictions at high temperature, 
they contradict at low temperature,
indicating that the logarithmic divergence of $S/T$ does not necessarily capture the intrinsic quantum criticality.
Rather, the contrasting behavior found at sufficiently low temperature 
would be crucial for understanding the QCP.
% Thus, for both experiment and theory, insight into the NFL properties at sufficiently low temperature is indispensable to unravel the nature of the quantum criticality.
% In a model of charge carriers on a 3D-Fermi surface scattered by 2D-AF
% fluctuations, transport properties near the QCP are found to be dominated
% by "hot spots``, points on the Fermi surface connected by the ordering wavevector. 
% In this case, calculations show that $C_{\rm el}/T\propto-\ln T$ and $S/T\propto-\ln T$.
% At first glance, this prediction seems consistent with the observed logarithmic increase of $-S/T$ at $T\geq$ 0.2 K.
As further increasing the fields, $-S(T)/T$ turns to increase below 60 mK at 1.5 T
and shows a continuous increase down to the lowest temperature at 2.0 T.
Above 3 T, the saturation behavior eventually appears after taking broad maxima,
which roughly coincide with the crossover temperature to the FL state~\cite{gegenwart}.
These observations are qualitatively consistent with the results of S. Hartmann~\cite{hartmann} in which the magnetic field is 
applied perpendicular to the $c$-axis.
\begin{figure}[t]
\includegraphics[scale =0.58]{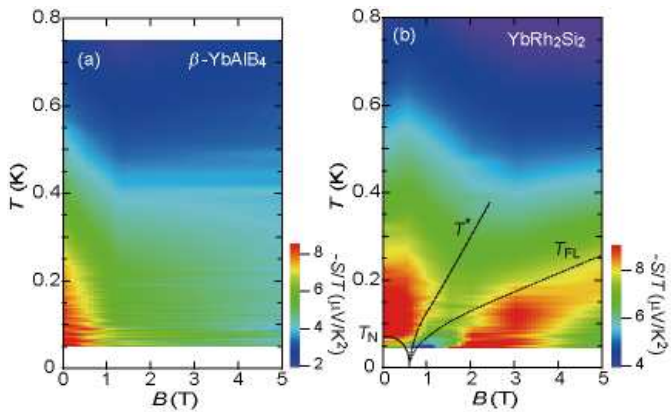}
\caption{\label{fig.2} (color online). Contour plots of  the Seebeck coefficient $-S/T$ for (a) $\beta$-YbAlB$_{4}$ and (b) YbRh$_2$Si$_2$
in the field-temperature plane.
$T_{\rm N}$, $T_{\rm FL}$, and $T^*$ represent the Neel ordering temperature, the crossover temperature to the Fermi liquid state, and
the crossover line where the Hall coefficient shows a rapid change, respectively~\cite{paschen}.}
\end{figure}

To clearly demonstrate the contrasting behavior of $S/T$ near the QCP,
we depict the contour plots of $-S(T)/T$ in the $B$-$T$ plane for (a) $\beta$-YbAlB$_{4}$ and (b) YbRh$_2$Si$_2$ in Fig.~\ref{fig.2}, respectively.
One immediately notices that in the vicinity of the zero-field QCP ($B_{\rm c}$ = 0 T), $-S(T)/T$ for $\beta$-YbAlB$_{4}$ is very large,
while the one for YbRh$_2$Si$_2$ is strongly suppressed around the field-induced QCP ($B_{\rm c}$ = 0.66 T).
% This might capture the striking difference of the quantum criticality of the two systems.
% We should emphasize that these constasting behavior of the Seebeck coefficient around the QCP indicate
% the distinct nature of quantum criticality in these two Yb-based heavy ferimion compounds.
In addition, we find two enhanced regions in the plot of YbRh$_2$Si$_2$; 
one is a zero-field region above $T_{\rm N}$ and the other is a zero-temperature region around 2 T.
% as has been found in the previous report~\cite{hartmann}, although a field scale is 10 times smaller than that of our case
% because of difference of the field direction.
Moreover, the $T^*$ line runs through between two enhanced regions.
% For YRS, we also found the two enhanced region of $-S(T)/T$; one is around above $T_{\rm N}$ and the other is around 2 T
% and the $T^*$ line seems run through between the two enhanced regions.
% At this moment, an origin of the enhancement is not clear, 
It seems that the evolution of $-S/T$ towards the QCP due to increasing QP mass
is unexpectedly suppressed around the QCP.
% , and hence one hot region is divided into two parts.
An interpretation of  this suppression will be discussed later.
On the other hand, 
the fact that the large $-S/T$ survives up to as high as 3 T, away from the QCP, implies
an additional source for the enhancement.
The field-induced FM fluctuations detected by the NMR measurements~\cite{ishida} could be a candidate.
We also mention that the persistent evolution of $-S/T$ in the low-field side of $T^*$ line, 
which amounts to as much as the one in the FL state,
seems contradict with the Kondo destruction scenario predicting the presence of the small Fermi surface~\cite{si,coleman}.
% predicted by the Kondo breakdown scenario~\cite{si,coleman}.
% Anyhow, we stress that the Seebeck coefficient unveils the several intriguing features associated with
% the low-energy QPs excitations in $\beta$-YbAlB$_{4}$ and YbRh$_2$Si$_2$, which have never been accessed by other probes. 
% According to the NMR experiments revealing the field-induced ferromagnetic fluctuations, the enhancement of $-S(T)/T$
% around around 2 T may attribute to this.
% The origin of the enhancement around $T_{\rm N}$ is unclear at this moment, but it may relate to the precursor of the AF order.
% We note that these characteristic features of the thermopower of YRS shares striking similarities with CeCoIn$_5$ as
% shown in Fig.~\ref{fig.4} (c).
% Namely, the contour plot of CeCoIn$_5$ has a suppressed region which terminates around the field-induced QCP, $B_c\sim$ 5.3 T. 
% This suggests that the 
\begin{figure}[t]
\begin{center}
\includegraphics[scale =0.45]{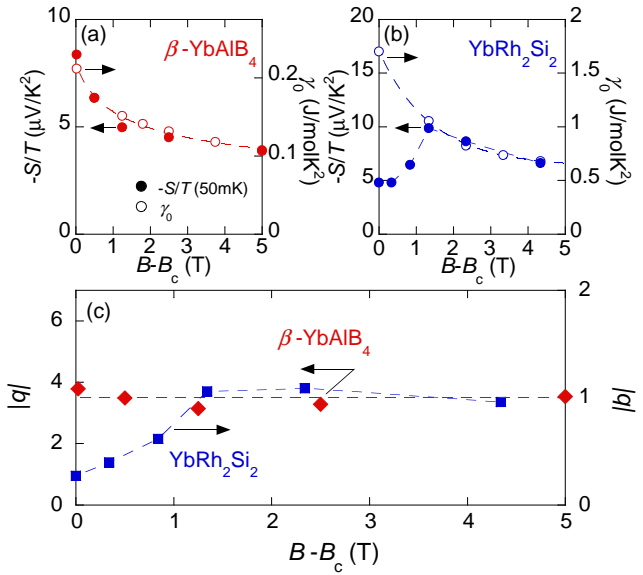}
\end{center}
\vspace{-0.5cm}
\caption{\label{fig.3} (color online). Field dependence of the Seebeck coefficient $-S/T$ at 50 mK (left axis) and
the electronic specific heat coefficient $\gamma_0$ (right axis)~\cite{nakatsuji,gegenwart}
for (a) $\beta$-YbAlB$_{4}$ and (b) YbRh$_2$Si$_2$, respectively.
(c) The dimensionless ratio $|q|$ vs $B-B_{\rm c}$ plot for ${\rm \beta}$-YbAlB$_{4}$ (left axis) and YbRh$_2$Si$_2$ (right axis).
The dashed lines are guide to the eye.}
\end{figure}

Let us further discuss the contrasting behavior of $S/T$ 
% as a clue to discriminate the nature of quantum criticality 
by examining the dimensionless ratio $q=(S/T\gamma_0)N_{\rm AV}e$.
Figure~\ref{fig.3} displays the field dependence of $-S(B)/T$ at 50 mK and $\gamma_0(B)$~\cite{nakatsuji,gegenwart,aaa}
for (a) $\beta$-YbAlB$_{4}$ and (b) YbRh$_2$Si$_2$, respectively.
% Here, $\gamma_0$ right at the QCP is taken from the temperature dependence of the electronic specific heat, 
% $C_{\rm el}(T)/T\equiv\gamma_0$ at 50 mK~\cite{nakatsuji,oeschler,aaa}.
% (For $\beta$-YbAlB$_{4}$, $C_{\rm el}(T)/T$ is estimated from the extrapolation by assuming $C_{\rm el}(T)/T\propto-\ln T$.)
It is clearly seen that $\gamma_0(B)$ diverges as $B\rightarrow B_{\rm c}$ in both systems,
indicating a divergence of the QP mass on approaching the QCP~\cite{nakatsuji,gegenwart}.
As expected from Eq. (1), $-S(B)/T$ for $\beta$-YbAlB$_{4}$ rapidly grows 
in parallel with $\gamma_0(B)$.
On the contrary, although $-S(B)/T$ trucks $\gamma_0(B)$ at high field in YbRh$_2$Si$_2$, 
it turns to decrease  as $B\rightarrow B_{\rm c}$.
Consequently, $|q|$ appears to be field-independent $|q|\sim 4$ in $\beta$-YbAlB$_4$, 
while the one of YbRh$_2$Si$_2$ shows a considerable decrease towards the QCP from a nearly constant value ($|q|\sim 1$) at high field (Fig.~\ref{fig.3}(c)).
Therefore, the semi-universality of $q$ is held in $\beta$-YbAlB$_4$ even at the QCP, while
it becomes invalid for YbRh$_2$Si$_2$ on approaching its QCP.
This striking difference of $q$ seems to point to distinct nature of the quantum criticality in these systems.

It has been pointed out that
the semi-universality of $q$ is satisfied when the whole Fermi surface (FS) equally contributes to $S/T$ and $\gamma_0$~\cite{miyake}.
This is a case for the QCP with $Q$ = 0 fluctuations
where the QP mass is enhanced on the entire region of the FS ( the whole FS is ``hot"), leading to the uniform enhancement of $S/T$ and $\gamma_0$.
% As we mentioned above, the Seebeck coefficient measures the delocalized quasiparticle (QP) density of state at the Fermi level.
However, the semi-universality is invalid for the case of AF ($Q\neq$ 0) QCP
because the mass enhancement occurs on some parts of the FS,
``hot regions" connected by the AF ordering wavevector $Q$.
% It has been pointed out that the hot electrons less participate in transport than the cold ones
% because the QP lifetime of the hot carriers is less than that of the cold carriers
% due to enhanced scattering with AF fluctuations.
In this case, since the transport is dominated by the cold electrons, $S/T$ is diminished due to a reduction of the hot electrons contribution, 
while $\gamma_0$ is enhanced because
it always has a contribution from the whole FS.
% By contrast, the specific heat always has the contribution from the whole FS.
% an emergence of ``hot spots", points on the Fermi surface connected by the ordering wavevector.
This leads to imbalance of $S/T$ and $\gamma_0$, and results in the reduction of $|q|$ from order of unity.
In fact, a considerable decrease of $|q|$ near the QCP was also found in CeCoIn$_5$~\cite{izawa},
for which the AF (SDW) scenario well accounts for the QCP.
% On the other hand, in the case of the QCP with fluctuations at $Q$ = 0, 
% the whole FSs uniformly contribute to $S/T$ and $\gamma_0$, and thus $|q|$ retains $\sim$1.
% since whole FSs are uniformly ``hot" without preferred spatial correlations.
According to this interpretation, the unchanged ($|q|\sim 4$) and the diminishing $|q|$ ($|q|< 1$) towards the QCP suggest
dominant $Q$ = 0 and finite $Q$ fluctuations in $\beta$-YbAlB$_4$ and YbRh$_2$Si$_2$, respectively.
Apparently, this finding is at odds with the theoretical predictions which attempt to understand the quantum criticality of the two systems 
in a unified way~\cite{imada,watanabe}.
% In addition, with respect to the $Q = 0$ fluctuations proposed in the local quantum criticality~\cite{si,coleman} and the quantum valence criticality~\cite{watanabe},
% the dominant $Q$ = 0 fluctuations in YbRh$_2$Si$_2$ suggested form $|q|<1$ seem to be incompatible with these scenarios, 
% while which remain as candidates for $\beta$-YbAlB$_4$.}
\begin{figure}[t]
\begin{center}
\includegraphics[scale =0.45]{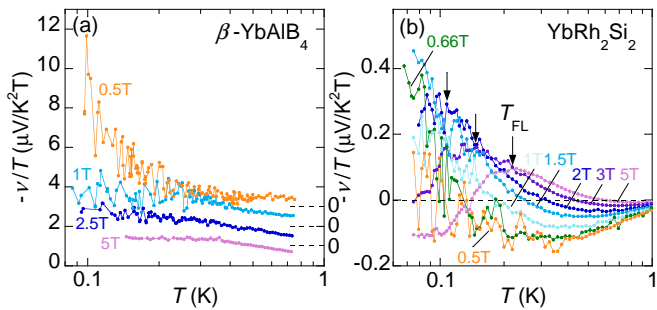}
\end{center}
\vspace{-0.5cm}
\caption{\label{fig.4} (color online). Temperature dependence of the Nernst coefficient $-\nu(T)/T$ 
for (a) $\beta$-YbAlB$_{4}$ and (b) YbRh$_2$Si$_2$ under several magnetic fields, respectively.
The $-\nu(T)/T$ curves for $\beta$-YbAlB$_{4}$ at $B\leq2.5$ T are sifted for clarity.}
\end{figure}

In contrast to the Seebeck coefficient, the Nernst coefficient $\nu/T$ is found to diverge in the vicinity of the QCP in both systems.
Figure~\ref{fig.4}(a) and \ref{fig.4}(b) represent
$-\nu(T)/T$ as a function of temperature for $\beta$-YbAlB$_4$ and YbRh$_2$Si$_2$, respectively, under several fields.
In $\beta$-YbAlB$_4$, we find a diverging increase of $-\nu(T)/T$, reaching as much as 7 $\mu$V/K$^2$T at 0.5 T,
while the enhancement is rapidly suppressed by the field of $B\geq$ 1 T.
We note that $-\nu(T)/T$ is very large, being three order of magnitude larger than that of simple metals~\cite{behnia_nernst}.
Since  the Nernst coefficient is expressed as the energy derivative of the Hall angle, 
$\nu/T\propto(\partial{\rm tan}\theta_{\rm H}/\partial\epsilon)_{\epsilon=\epsilon_{\rm F}}$,
$\nu/T$ is very small if tan$\theta_{\rm H}$ is only weakly dependent on energy, which is a good
approximation in simple metals~\cite{behnia_nernst}.
On the other hand, in the case of the energy-dependent tan$\theta_{\rm H}$, $\nu/T$ can be obtained from the
following simple expression, valid as $T\rightarrow0$~\cite{behnia_nernst}:
$\frac{\nu}{T}\simeq\frac{\pi^2}{3}\frac{k_{\rm B}^2}{e}\frac{\mu}{\epsilon_{\rm F}}$,
where $\mu$ represents the carrier mobility.
% and $\nu$ can be of either sign with no direct relation to carrier type.
Therefore, the divergence of $-\nu(T)/T$ is interpreted as a consequence of vanishing of $\epsilon_{\rm F}$,
defined as the characteristic energy scale of the FL.
A similar diverging enhancement of $-\nu(T)/T$ is also observed under the critical field ($B_{\rm c}$ = 0.66 T) in YbRh$_2$Si$_2$ (Fig.~\ref{fig.4}(b)), 
although the behaviors of $-\nu(T)/T$ is rather complicated including a sign change.
By further applying the fields, the diverging behavior is gradually suppressed and peaks are formed above 2 T (downward arrows).
These peak temperatures correspond to $T_{\rm FL}$ in Ref.~\cite{gegenwart}, indicating that the FL is restored below the peak temperatures.
% Indeed, the contour plot of $-\nu(T)/T$ shown in Fig.~\ref{fig.4}(c) denotes that $-\nu(T)/T$ is large along the crossover line of $T_{\rm FL}$.
% which allows for an accurate determination of the onset of FL behavior.
% In addition, the diverging enhancement of $-\nu(T)/T$ clearly 
% demonstrates the pronounced vanishing of $\epsilon_{\rm F}$ towards the QCP in both systems.
% without need for
% any subtraction of an offset in contrary to other probes, such as the specific heat and the resistivity.
% This indicates that the Nernst coefficient is a sensitive probe of the energy scale of the FL, and hence useful to capture the quantum criticality.
% Notably, a similar enhancement of $\nu/T$ associated with the QCP was also observed in CeCoIn$_5$~\cite{izawa}.
% We also note that the sign reversal of $-\nu(T)/T$ occurs below and above the enhanced region.
The sigh change of $-\nu(T)/T$ suggests an existence of two FSs with dominant carriers of opposite signs 
in accordance with the band calculations~\cite{wigger},
which may result in the relatively small magnitude of $-\nu(T)/T$ compared with $\beta$-YbAlB$_4$ 
due to canceling of opposite sign contributions to the Nernst signal.
% On the other hand, the fact that one of the sign reversal takes place close to the $T^*$ line is reminiscence of
% a deformation of the electronic structure as suggested by the sudden change of the Hall effect~\cite{paschen},
% but its relevance to the Kondo breakdown is yet to be clarified.
% A further theoretical investigation of this issue is required.

In summary,
we present the low-temperature thermoelectric coefficient measurements of $\beta$-YbAlB$_4$ and YbRh$_2$Si$_2$.
Both systems display the diverging enhancement of the Nernst coefficient on approaching their QCPs, indicative of the
vanishing of the Fermi energy.
On the other hand, the significant deviation from the semi-universal ratio $q$ of the Seebeck coefficient and the electronic specific heat
is found in YbRh$_2$Si$_2$ in the vicinity of the QCP, while the universality is sustained in $\beta$-YbAlB$_4$.
This striking difference suggests distinct type of fluctuations underlying the quantum criticality; $Q$ = 0 and finite $Q$ fluctuations for $\beta$-YbAlB$_4$
and YbRh$_2$Si$_2$, respectively.
Further experiments on other quantum critical materials in terms of $q$ 
would enable systematic clarification of the quantum criticality. 

% We have found a apparent deviation from the universal value of the dimensionless ratio
% of the Seebeck coefficient and the electronic specific heat in YRS,
% while 
% In summary, we have demonstrated that the two representative Yb-based unconventional 
% quantum critical materials $\beta$-YbAlB$_4$ and YbRh$_2$Si$_2$,
% belong the distinct class of the quantum criticality 
% from the dimensionless ratio of the Seebeck coefficient and the electronic specific heat.

% We thank K. Behnia, H. Kusunose, K. Miyake, and S. Watanabe for discussions.
This work is supported by Grants-in-Aids (Nos. 23340099, 23740263, 21684019) 
for Scientific Research from the Japanese Society for the Promotion of Science,
a Grant-in-Aid for Scientific Research on
Innovative Areas ``Heavy Electrons" (Nos. 20102006, 23102706) of
the Ministry of Education, Culture, Sports, Science, and Technology (MEXT),
and a Grant-in-Aid for the Global COE Program from the MEXT through the Nanoscience and
Quantum Physics Project of the Tokyo Institute of Technology.

\bibliography{seebeck}

\begin{thebibliography}{30}
\expandafter\ifx\csname natexlab\endcsname\relax\def\natexlab#1{#1}\fi
\expandafter\ifx\csname bibnamefont\endcsname\relax
  \def\bibnamefont#1{#1}\fi
\expandafter\ifx\csname bibfnamefont\endcsname\relax
  \def\bibfnamefont#1{#1}\fi
\expandafter\ifx\csname citenamefont\endcsname\relax
  \def\citenamefont#1{#1}\fi
\expandafter\ifx\csname url\endcsname\relax
  \def\url#1{\texttt{#1}}\fi
\expandafter\ifx\csname urlprefix\endcsname\relax\def\urlprefix{URL }\fi
\providecommand{\bibinfo}[2]{#2}
\providecommand{\eprint}[2][]{\url{#2}}

\bibitem[{\citenamefont{Gegenwart et~al.}(2008)\citenamefont{Gegenwart, Si, and
  Steglich}}]{gegenwart_review}
\bibinfo{author}{\bibfnamefont{P.}~\bibnamefont{Gegenwart}},
  \bibinfo{author}{\bibfnamefont{Q.}~\bibnamefont{Si}}, \bibnamefont{and}
  \bibinfo{author}{\bibfnamefont{F.}~\bibnamefont{Steglich}},
  \bibinfo{journal}{Nature Phys.} \textbf{\bibinfo{volume}{4}},
  \bibinfo{pages}{186} (\bibinfo{year}{2008}).

\bibitem[{\citenamefont{Moriya and Takimoto}(1995)}]{moriya}
\bibinfo{author}{\bibfnamefont{T.}~\bibnamefont{Moriya}} \bibnamefont{and}
  \bibinfo{author}{\bibfnamefont{T.}~\bibnamefont{Takimoto}},
  \bibinfo{journal}{J. Phys. Soc. Jpn.} \textbf{\bibinfo{volume}{64}},
  \bibinfo{pages}{960} (\bibinfo{year}{1995}).

\bibitem[{\citenamefont{Hertz}(1976)}]{hertz}
\bibinfo{author}{\bibfnamefont{J.~A.} \bibnamefont{Hertz}},
  \bibinfo{journal}{Phys. Rev. B} \textbf{\bibinfo{volume}{14}},
  \bibinfo{pages}{1165} (\bibinfo{year}{1976}).

\bibitem[{\citenamefont{Millis}(1993)}]{millis}
\bibinfo{author}{\bibfnamefont{A.~J.} \bibnamefont{Millis}},
  \bibinfo{journal}{Phys. Rev. B} \textbf{\bibinfo{volume}{7183}},
  \bibinfo{pages}{48} (\bibinfo{year}{1993}).

\bibitem[{\citenamefont{Knafo et~al.}(2009)\citenamefont{Knafo, Raymond, Lejay,
  and Flouquet}}]{knafo}
\bibinfo{author}{\bibfnamefont{W.}~\bibnamefont{Knafo}},
  \bibinfo{author}{\bibfnamefont{S.}~\bibnamefont{Raymond}},
  \bibinfo{author}{\bibfnamefont{P.}~\bibnamefont{Lejay}}, \bibnamefont{and}
  \bibinfo{author}{\bibfnamefont{J.}~\bibnamefont{Flouquet}},
  \bibinfo{journal}{Nature Phys.} \textbf{\bibinfo{volume}{5}},
  \bibinfo{pages}{753} (\bibinfo{year}{2009}).

\bibitem[{\citenamefont{v.~L$\ddot{\rm o}$hneysen et~al.}(1994)}]{lohneysen}
\bibinfo{author}{\bibfnamefont{H.}~\bibnamefont{v.~L$\ddot{\rm o}$hneysen}}
  \bibnamefont{et~al.}, \bibinfo{journal}{Phys. Rev. Lett.}
  \textbf{\bibinfo{volume}{72}}, \bibinfo{pages}{3262} (\bibinfo{year}{1994}).

\bibitem[{\citenamefont{Gegenwart et~al.}(2002)}]{gegenwart}
\bibinfo{author}{\bibfnamefont{P.}~\bibnamefont{Gegenwart}}
  \bibnamefont{et~al.}, \bibinfo{journal}{Phys. Rev. Lett.}
  \textbf{\bibinfo{volume}{89}}, \bibinfo{pages}{056402}
  (\bibinfo{year}{2002}).

\bibitem[{\citenamefont{Nakatsuji et~al.}(2008)}]{nakatsuji}
\bibinfo{author}{\bibfnamefont{S.}~\bibnamefont{Nakatsuji}}
  \bibnamefont{et~al.}, \bibinfo{journal}{Nature Phys.}
  \textbf{\bibinfo{volume}{4}}, \bibinfo{pages}{603} (\bibinfo{year}{2008}).

\bibitem[{\citenamefont{Izawa et~al.}(2007)}]{izawa}
\bibinfo{author}{\bibfnamefont{K.}~\bibnamefont{Izawa}} \bibnamefont{et~al.},
  \bibinfo{journal}{Phys. Rev. Lett.} \textbf{\bibinfo{volume}{99}},
  \bibinfo{pages}{147005} (\bibinfo{year}{2007}).

\bibitem[{\citenamefont{Hartmann et~al.}(2010)}]{hartmann}
\bibinfo{author}{\bibfnamefont{S.}~\bibnamefont{Hartmann}}
  \bibnamefont{et~al.}, \bibinfo{journal}{Phys. Rev. Lett.}
  \textbf{\bibinfo{volume}{104}}, \bibinfo{pages}{096401}
  (\bibinfo{year}{2010}).

\bibitem[{\citenamefont{Matsumoto et~al.}(2011)}]{matsumoto}
\bibinfo{author}{\bibfnamefont{Y.}~\bibnamefont{Matsumoto}}
  \bibnamefont{et~al.}, \bibinfo{journal}{Science}
  \textbf{\bibinfo{volume}{21}}, \bibinfo{pages}{316} (\bibinfo{year}{2011}).

\bibitem[{\citenamefont{Okawa et~al.}(2010)}]{okawa}
\bibinfo{author}{\bibfnamefont{M.}~\bibnamefont{Okawa}} \bibnamefont{et~al.},
  \bibinfo{journal}{Phys. Rev. Lett.} \textbf{\bibinfo{volume}{104}},
  \bibinfo{pages}{247201} (\bibinfo{year}{2010}).

\bibitem[{\citenamefont{Custers et~al.}(2003)}]{custers}
\bibinfo{author}{\bibfnamefont{J.}~\bibnamefont{Custers}} \bibnamefont{et~al.},
  \bibinfo{journal}{Nature} \textbf{\bibinfo{volume}{423}},
  \bibinfo{pages}{524} (\bibinfo{year}{2003}).

\bibitem[{\citenamefont{Paschen et~al.}(2004)}]{paschen}
\bibinfo{author}{\bibfnamefont{S.}~\bibnamefont{Paschen}} \bibnamefont{et~al.},
  \bibinfo{journal}{Nature} \textbf{\bibinfo{volume}{432}},
  \bibinfo{pages}{881} (\bibinfo{year}{2004}).

\bibitem[{\citenamefont{Custers et~al.}(2010)}]{custers2}
\bibinfo{author}{\bibfnamefont{J.}~\bibnamefont{Custers}} \bibnamefont{et~al.},
  \bibinfo{journal}{Phys. Rev. Lett.} \textbf{\bibinfo{volume}{186402}},
  \bibinfo{pages}{104} (\bibinfo{year}{2010}).

\bibitem[{\citenamefont{Misawa et~al.}(2009)\citenamefont{Misawa, Yamaji, and
  Imada}}]{imada}
\bibinfo{author}{\bibfnamefont{T.}~\bibnamefont{Misawa}},
  \bibinfo{author}{\bibfnamefont{Y.}~\bibnamefont{Yamaji}}, \bibnamefont{and}
  \bibinfo{author}{\bibfnamefont{M.}~\bibnamefont{Imada}}, \bibinfo{journal}{J.
  Phys. Soc. Jpn.} \textbf{\bibinfo{volume}{78}}, \bibinfo{pages}{084707}
  (\bibinfo{year}{2009}).

\bibitem[{\citenamefont{Watanabe and Miyake}(2010)}]{watanabe}
\bibinfo{author}{\bibfnamefont{S.}~\bibnamefont{Watanabe}} \bibnamefont{and}
  \bibinfo{author}{\bibfnamefont{K.}~\bibnamefont{Miyake}},
  \bibinfo{journal}{Phys. Rev. Lett.} \textbf{\bibinfo{volume}{105}},
  \bibinfo{pages}{186403} (\bibinfo{year}{2010}).

\bibitem[{\citenamefont{Macaluso et~al.}(2007)}]{robin}
\bibinfo{author}{\bibfnamefont{R.~T.} \bibnamefont{Macaluso}}
  \bibnamefont{et~al.}, \bibinfo{journal}{Chem. Mater.}
  \textbf{\bibinfo{volume}{19}}, \bibinfo{pages}{1918} (\bibinfo{year}{2007}).

\bibitem[{\citenamefont{Knebel et~al.}(2006)}]{knebel}
\bibinfo{author}{\bibfnamefont{G.}~\bibnamefont{Knebel}} \bibnamefont{et~al.},
  \bibinfo{journal}{J. Phys. Soc. Jpn.} \textbf{\bibinfo{volume}{75}},
  \bibinfo{pages}{114709} (\bibinfo{year}{2006}).

\bibitem[{\citenamefont{Behnia et~al.}(2004)\citenamefont{Behnia, Jaccard, and
  Flouquet}}]{behnia_seebeck}
\bibinfo{author}{\bibfnamefont{K.}~\bibnamefont{Behnia}},
  \bibinfo{author}{\bibfnamefont{D.}~\bibnamefont{Jaccard}}, \bibnamefont{and}
  \bibinfo{author}{\bibfnamefont{J.}~\bibnamefont{Flouquet}},
  \bibinfo{journal}{J. Phys.: Condens. Matter} \textbf{\bibinfo{volume}{16}},
  \bibinfo{pages}{5187} (\bibinfo{year}{2004}).

\bibitem[{\citenamefont{Kuga et~al.}(2008)}]{kuga}
\bibinfo{author}{\bibfnamefont{K.}~\bibnamefont{Kuga}} \bibnamefont{et~al.},
  \bibinfo{journal}{Phys. Rev. Lett} \textbf{\bibinfo{volume}{101}},
  \bibinfo{pages}{137004} (\bibinfo{year}{2008}).

\bibitem[{\citenamefont{Paul and Kotliar}(2001)}]{kotliar}
\bibinfo{author}{\bibfnamefont{I.}~\bibnamefont{Paul}} \bibnamefont{and}
  \bibinfo{author}{\bibfnamefont{G.}~\bibnamefont{Kotliar}},
  \bibinfo{journal}{Phys. Rev. B} \textbf{\bibinfo{volume}{64}},
  \bibinfo{pages}{184414} (\bibinfo{year}{2001}).

\bibitem[{\citenamefont{Kim and P$\acute{\rm e}$pin}(2010)}]{pepin}
\bibinfo{author}{\bibfnamefont{K.-S.} \bibnamefont{Kim}} \bibnamefont{and}
  \bibinfo{author}{\bibfnamefont{C.}~\bibnamefont{P$\acute{\rm e}$pin}},
  \bibinfo{journal}{Phys. Rev. B} \textbf{\bibinfo{volume}{81}},
  \bibinfo{pages}{205108} (\bibinfo{year}{2010}).

\bibitem[{\citenamefont{Ishida et~al.}(2002)}]{ishida}
\bibinfo{author}{\bibfnamefont{K.}~\bibnamefont{Ishida}} \bibnamefont{et~al.},
  \bibinfo{journal}{Phys. Rev. Lett.} \textbf{\bibinfo{volume}{89}},
  \bibinfo{pages}{107202} (\bibinfo{year}{2002}).

\bibitem[{\citenamefont{Si et~al.}(2001)\citenamefont{Si, Rabello, Ingersent,
  and Smith}}]{si}
\bibinfo{author}{\bibfnamefont{Q.}~\bibnamefont{Si}},
  \bibinfo{author}{\bibfnamefont{S.}~\bibnamefont{Rabello}},
  \bibinfo{author}{\bibfnamefont{K.}~\bibnamefont{Ingersent}},
  \bibnamefont{and} \bibinfo{author}{\bibfnamefont{J.~L.} \bibnamefont{Smith}},
  \bibinfo{journal}{Nature} \textbf{\bibinfo{volume}{413}},
  \bibinfo{pages}{804} (\bibinfo{year}{2001}).

\bibitem[{\citenamefont{Coleman et~al.}(2001)\citenamefont{Coleman,
  P$\acute{\rm e}$pin, Si, and Ramazashvili}}]{coleman}
\bibinfo{author}{\bibfnamefont{P.}~\bibnamefont{Coleman}},
  \bibinfo{author}{\bibfnamefont{C.}~\bibnamefont{P$\acute{\rm e}$pin}},
  \bibinfo{author}{\bibfnamefont{Q.}~\bibnamefont{Si}}, \bibnamefont{and}
  \bibinfo{author}{\bibfnamefont{R.}~\bibnamefont{Ramazashvili}},
  \bibinfo{journal}{J. Phys.: Condens. Matter} \textbf{\bibinfo{volume}{13}},
  \bibinfo{pages}{R723} (\bibinfo{year}{2001}).

\bibitem[{aaa()}]{aaa}
\bibinfo{note}{The value of $\gamma_0$ at the QCP is taken from the temperature
  dependence of the electronic specific heat, $C_{\rm el}(T)/T\equiv\gamma_0$
  at 50 mK~\cite{nakatsuji,custers}. In addition, for $\beta$-YbAlB$_{4}$,
  $C_{\rm el}(T)/T$ at 50 mK is estimated from the extrapolation by assuming
  $C_{\rm el}(T)/T\propto-\ln T$.}

\bibitem[{\citenamefont{Miyake and Kohno}(2005)}]{miyake}
\bibinfo{author}{\bibfnamefont{K.}~\bibnamefont{Miyake}} \bibnamefont{and}
  \bibinfo{author}{\bibfnamefont{H.}~\bibnamefont{Kohno}}, \bibinfo{journal}{J.
  Phys. Soc. Jpn.} \textbf{\bibinfo{volume}{74}}, \bibinfo{pages}{254}
  (\bibinfo{year}{2005}).

\bibitem[{\citenamefont{Behnia}(2009)}]{behnia_nernst}
\bibinfo{author}{\bibfnamefont{K.}~\bibnamefont{Behnia}}, \bibinfo{journal}{J.
  Phys.: Condens. Matter} \textbf{\bibinfo{volume}{21}},
  \bibinfo{pages}{113101} (\bibinfo{year}{2009}).

\bibitem[{\citenamefont{Wigger et~al.}(2007)}]{wigger}
\bibinfo{author}{\bibfnamefont{G.~A.} \bibnamefont{Wigger}}
  \bibnamefont{et~al.}, \bibinfo{journal}{Phys. Rev. B}
  \textbf{\bibinfo{volume}{76}}, \bibinfo{pages}{035106}
  (\bibinfo{year}{2007}).

\end{thebibliography}
\end{document}